\documentclass[aps,prl,showpacs,twocolumn]{revtex4}
\usepackage{graphics}
\usepackage[dvips]{color}
\usepackage[T1]{fontenc}
\usepackage[latin1]{inputenc}
\usepackage{graphicx}
\usepackage{amssymb}
\usepackage{rotating}
\usepackage{epsfig}

\input epsf.sty

\newcommand{\sts}{\scriptsize}

\newcommand{\mb}{\mbox}

\newcommand{\ie}{{\em i.e.}} 

\newcommand{\eg}{{\em e.g.}}

\begin{document}

\title{A Coarse-Grained Lattice Model for Molecular Recognition}
\author{Hans Behringer, Andreas Degenhard, Friederike Schmid}
\affiliation{Fakult\"at f\"ur Physik, Universit\"at Bielefeld, D -- 33615
Bielefeld, Germany}

\begin{abstract}
  We present a simple model which allows to investigate
  equilibrium aspects of molecular recognition between rigid
  biomolecules on a generic level. Using a two-stage approach, which 
  consists of a design and a testing step, the role of cooperativity 
  and of varying bond strength in molecular recognition is investigated. 
  Cooperativity is found to enhance selectivity. In complexes which require 
  a high binding flexibility a small number of strong bonds seems to be
  favored compared to a situation with many but weak bonds. 
\end{abstract}

\pacs{87.15.-v, 87.15.Aa, 89.20.-a}
\maketitle
Living organisms could not function without the ability of 
biomolecules to specifically recognize each other
\cite{Alberts_1994,Kleanthous_2000}.  Molecular recognition can be
viewed as the ability of a biomolecule to interact preferentially with
a particular target molecule among a vast variety of different but
structurally similar rival molecules. Recognition processes are
governed by an interplay of non-covalent interactions, in particular,
hydrophobic interactions and hydrogen bonds.
Such non-covalent bonds have typical energies of 1-2 kcal/mole (the relatively
strong hydrogen bonds may contribute up to 8-10 kcal/mole)  and are
therefore only slightly stronger than the thermal energy $k_{\sts
  \mb{B}}T_{\sts \mb{Room}} \simeq 0.62$ kcal/mole at physiological
conditions. Biomolecular recognition is thus only achieved if a large 
number of functional groups on the two partner molecules match precisely. 
This observation has lead to a ``key-lock'' picture: Two biomolecules 
recognize each other if their shapes at the recognition site and/or 
the interactions between the residues in contact are 
largely complementary~\cite{pauling_1940}.

In the present Letter, we introduce a coarse-grained approach which allows 
to investigate this ``principle of complementarity'' on a very general level, 
and use it to study the role of different factors for the selectivity 
of interactions between biomolecule surfaces. Specifically, we analyze 
two elements that have been discussed in the literature: the cooperativity, 
and the interplay of interaction strengths. We will show that our model 
can help to understand some of the features of real protein-protein 
interfaces.

Previous theoretical studies have mostly dealt with the adsorption of
heteropolymers on random and structured surfaces
\cite{Chakraborty_2001,Polotsky_2004a,Bogner_2004}.
Some works have adapted the random energy model from the theory of 
disordered systems to the problem of biomolecular 
binding~\cite{Janin_1986,Wang_2003}.
In contrast, in the present approach, we consider explicitly systems of 
two interacting, rigid, heterogeneous surfaces. This is motivated by 
some basic findings about the biochemical structure of the 
recognition site, \ie, the contact interface between recognizing proteins. 
In recent years the structural properties of proteins at the 
recognition site has been clarified~\cite{Jones_1996,
LoConte_1999,Kleanthous_2000}. Although different protein-protein
complexes may differ considerably, a general picture of a standard
recognition site containing approximately 30 residues,
with a total size of 1200-2000 \AA${}^2$ has emerged.
Apart from notable exceptions, the association of the proteins is
basically rigid, although minor rearrangements of amino acid
side-chains do occur \cite{Jones_1996, LoConte_1999}.

We describe the structure of the proteins at the contact interfaces by 
two sets of classical spin variables $\sigma = (\sigma_1,\ldots, \sigma_N)$ 
and $\theta= (\theta_1, \ldots,\theta_N)$, whose values specify 
the various types of residues. The set $\sigma$ characterizes the
structure of the recognition site on the target molecule, and $\theta$ that
on the probe molecule, \ie, the molecule that is supposed to recognize the 
target.  The position of site $i$ on the surfaces can be specified arbitrarily. 
For simplicity, we assume that the positions $i$ on both surfaces match,
and that the total number of contact residues is equal $N$ for both 
molecules. However, we take into account the possibility that the 
quality of the contact of two residues at position $i$ may vary, \eg, 
due to steric hindrances or varying relative alignment of polar moments,
caused by minor rearrangements of the amino acid side-chains.
This is modeled by an additional variable $S_i$, $i=1, \ldots, N$. 
The total interaction is thus described by a Hamiltonian
$\mathcal{H}(\sigma, \theta; S)$, which incorporates in a coarse-grained
way both the structural properties of the recognition site and the 
interaction between residues.

To study the recognition process between two biomolecules, we adopt a 
two-stage approach. We take the structure of the target recognition site,
$\sigma^{(0)} = (\sigma_1^{(0)}, \ldots, \sigma_N^{(0)})$, to be given.
In the first step, the probe ``learns'' the target structure at a given 
``design temperature'' $1/\beta_{\mb{\tiny D}}$. One obtains an ensemble 
of probe molecules with structures $\theta$ distributed according to a 
probability $P(\theta | \sigma^{(0)}) = \frac{1}{Z_{\sts
    \mb{\tiny D}}} \sum_{S} \exp\left(-\beta_{\sts\mb{\tiny D}}
  \mathcal{H}(\sigma^{(0)}, \theta;S) \right)$, which depends on
the target structure. This first design step is introduced to mimic
the design in biotechnological applications or the evolution process
in nature. The parameter $\beta_{\sts\mb{\tiny D}}$ characterizes the
conditions under which the design has been carried out, \ie, it is
a Lagrange parameter which fixes the achieved average interaction energy.
A similar design procedure has been introduced in studies of protein 
folding \cite{Pande_2000} and the adsorption of polymers on structured 
surfaces \cite{Jayaraman_2005}.
In the second step, the recognition ability of the designed probe
ensemble is tested. To this end the probe molecules are exposed to
both the original target structure $\sigma^{(0)}$ and a competing 
(different) rival structure $\sigma^{(1)}$ at some temperature $1/{\beta}$, 
which in general differs from the design temperature $1/\beta_{\mb{\tiny D}}$. 
The thermal free energy $F(\theta|\sigma^{(\alpha)})$ for the
interaction between $\sigma^{(\alpha)}$ ($\alpha = 0,1$) and a probe $\theta$ 
is given by
$
F(\theta|\sigma^{(\alpha)}) 
= -\frac{1}{\beta} \ln \sum_{S} \exp\left(- \beta \mathcal{H}(\sigma^{(\alpha)},
  \theta;S) \right)
$.
Averaged over all probe molecules, we obtain $\langle F^{(\alpha)} \rangle 
= \sum_\theta F(\theta|\sigma^{(\alpha)})P(\theta|\sigma^{(0)})$.
The target is recognized if the average free energy difference
$\Delta F = \langle F^{(0)} \rangle - \langle F^{(1)} \rangle$ is negative,
\ie, probe molecules exposed to equal amounts of target and rival molecules
preferentially bind to the target. Note that our treatment does not 
account for kinetic effects, only equilibrium aspects are considered.

The association of the  proteins is accompanied by a reduction of the 
translational and rotational entropy. However, these additional entropic 
contributions to the free energy of association depend
only  weakly on the mass 
and shape of the rigid molecules, and can be considered, in a first 
approximation, to be of the same order for the association with the target 
and the rival molecule. Thus, these contributions cancel in the 
free energy difference. Similarly, contributions from the 
interaction with solvent molecules are also assumed to be of comparable size.

A modified HP-model can serve as a first example to illustrate 
this general description. In the HP-model, which was introduced originally
to study protein folding \cite{Dill_1985}, residues are distinguished
by their hydrophobicity only. Hydrophobic residues are represented by 
$\sigma_i,\theta_i = +1$, and polar residues by $\sigma_i,\theta_i = -1$. 
In addition, the variable $S_i$ describing the (geometric) quality of the contact can 
take on the values $\pm 1$ where $S_i = +1$ models a good contact and 
$S_i=-1$ a bad one. Only for good contacts does one get a contribution to 
the binding energy.  The Hamiltonian is then given by
\begin{equation}
\label{eq:HP-hamiltonian}
  \mathcal{H}(\sigma,\theta;S) 
=- \varepsilon \sum_{i}\frac{1+S_i}{2} \sigma_i\theta_i
\end{equation}  
where the sum extends over the $N$ positions of the residues of the
recognition site and $\varepsilon$ being the interaction
constant~\cite{footnote}. Note that a ``good'' contact can nevertheless
lead to an unfavorable energy contribution. For this simple model, the
different steps of the two-stage approach described above can be
worked out analytically.

First, we analyze the efficiency of the design step by inspecting the achieved 
complementarities (of interactions) of the designed probe molecules with the 
target molecule. To this end, we define a complementarity parameter 
$K = \sum_i \sigma^{(0)}_i\theta_i$ which ranges from $-N$ to $+N$, with 
$K$ close to $+N$ signaling a large complementarity of the recognition sites. 
The probability distribution $P(\theta|\sigma^{(0)})$ can be converted to a 
distribution $P(K)$ for the probability of having a complementarity $K$. 
Up to a normalization factor, it is given by
\begin{equation}
  P(K) \propto  {N \choose \frac{1}{2}(N+K)} \exp \left(
  \frac{\varepsilon\beta_{\mb{\tiny D}}}{2} K\right).
\end{equation}   
Its first moment $\left<K\right> = \sum_K K P(K) 
= N \tanh\left( {\varepsilon\beta_{\mb{\tiny D}}}/{2} \right)$
quantifies the quality of the design. For decreasing design 
temperatures $ 1/\beta_{\mb{\tiny D}}$ 
the average complementarity per site $\left< K \right>/N$ approaches
one, and thus the designed probe molecules are well optimized with
respect to the target. 

In the second step the association of the probe molecules with the
target and with a different rival molecule is compared. Introducing
the quantity $Q = \sum_i \sigma^{(0)}_i\sigma^{(1)}_i$ as a measure
for the similarity between the recognition sites of the target and the
rival molecules, the free energy difference per site can be expressed
in the form $\Delta F(Q)/N = -\frac{1}{2} \varepsilon
\tanh\left(\frac{\varepsilon \beta_{\mb{\tiny D}}}{2} \right)
(1-Q/N).$ $\Delta F/N$ is negative, if the rival and the target are
different and $Q$ is thus smaller than $N$. The probe molecule therefore
binds preferentially to the target molecule, and thus the target is
specifically recognized. The free energy difference increases with
decreasing similarity parameter $Q$.

After this introductory analysis of a simple system, we turn to consider
more complex models which allow to investigate the influence of different
factors on the specific recognition between surfaces. We begin with 
studying the role of cooperativity.

Systematic mutagenesis experiments have revealed that cooperativity
plays an important role in molecular recognition processes
\cite{Cera_1998}. Cooperativity in biological processes basically means 
that the interaction strength of two residues depends on the interactions 
in their neighborhood. 
Physically, this can be caused by a physical rearrangement of 
amino acid side-chains or a readjustment of polar moments as
a function of the local environment. In the simplified
language of our model, cooperativity thus means that the quality 
of a contact depends on the quality of the neighbor contacts.
This can be incorporated in the HP-model by the following extension:
\begin{equation}
\label{eq:HP-coop}
\mathcal{H}(\sigma, \theta;S) = 
- \varepsilon\sum_{i=1}^{N} \frac{1+S_i}{2} \sigma_i \theta_i - J
\sum_{\left<ij\right>} S_i S_j.
\end{equation}
The second sum accounts for the cooperative interaction and runs over 
neighbor residue positions $i$ and $j$. The interaction coefficient $J$ 
is positive for cooperative interactions and negative for anti-cooperativity.  
For $J > 0$, the cooperative term rewards additional contacts in the vicinity 
of a good contact between two residues. This leads to a better optimization 
of the side-chains and thus the complementarity between the probe and the 
target molecule is improved. Cooperativity is therefore expected to enhance 
the quality of the design step compared to an interaction without cooperativity.
Similarly, one expects an improved recognition specificity.

For non-zero, but finite values of $J$, the model can no longer be solved
analytically. Therefore, we calculated numerically the density of states 
for the interaction between two proteins as a function of the energy and the 
complementarity parameter using 
efficient modern Monte Carlo algorithms \cite{Hueller_2002}. The density of states $\Omega_J(K, E)$ for a fixed target structure 
$\sigma^{(0)}$ is the number of configurations $(\theta,S)$ that have energy 
$E=\mathcal{H}$ when interacting with the target, and a complementarity 
$K$ with the target recognition site.  
The probability distribution of the complementarity $K$
is then (up to a normalization constant) given by $ P_{\beta_{\mb{\tiny D}}}(K; J) \sim \sum_E
\Omega_J(K, E) \exp(-\beta_{\mb{\tiny D}}E)$.

\begin{figure}
\begin{center}
\includegraphics[scale=0.25,angle=0]{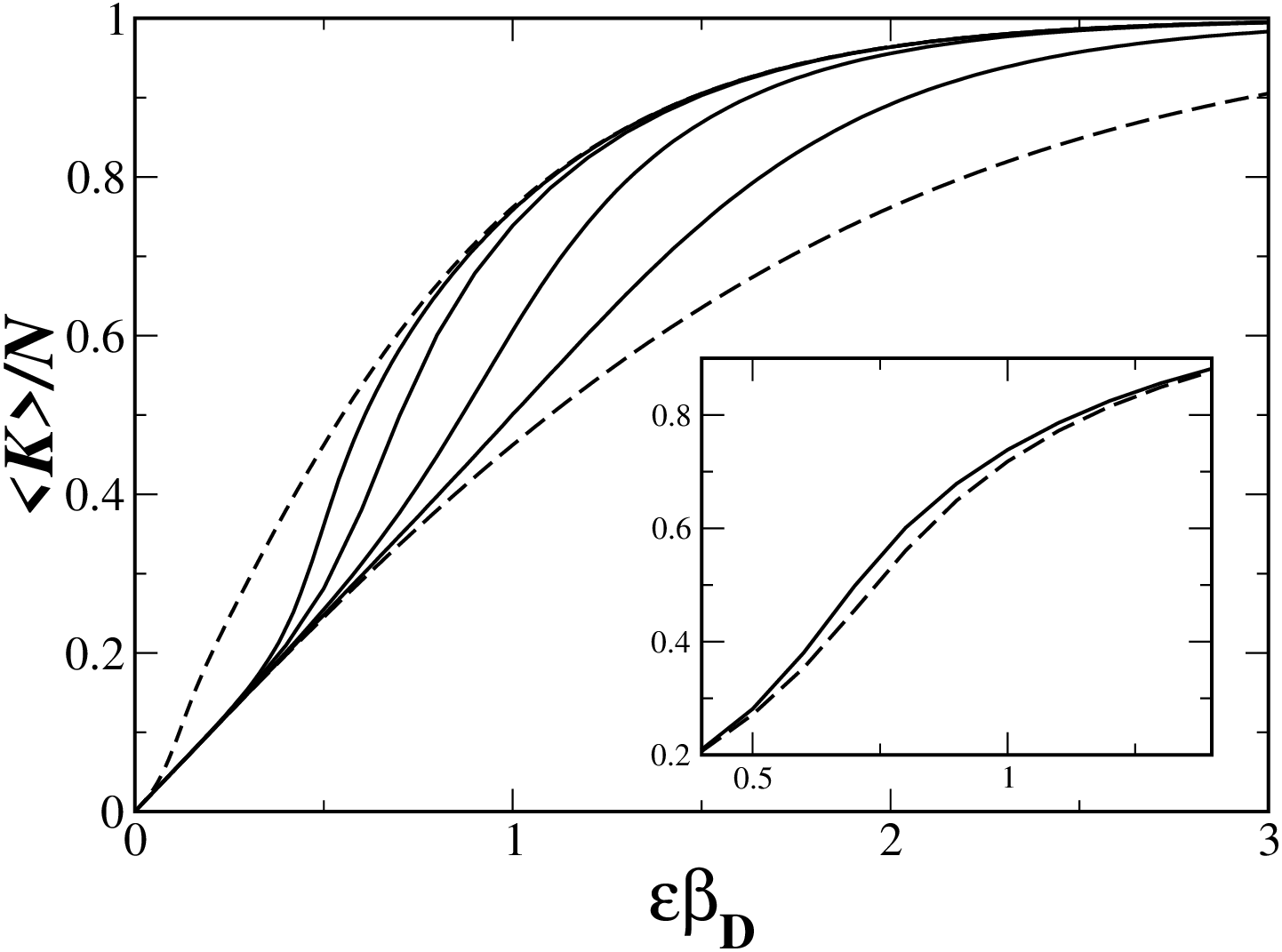}
\caption{\label{bild:coop_comp} 
  Average complementarity per site of the designed probe ensemble for different
  values of $J$. For the lower dashed curve $J=0$, 
  the upper dashed line represents the limit $J \to
  \infty$, which can be tackled analytically
  \cite{Behringer_etal_vor}. The curves in between from bottom up belong 
  to values 0.1, 0.25, 0.5, 0.75 of $J$ in units of $\varepsilon$. The
  inset shows  $\left<K\right>/N$ for $N=256 $ (full curve) and
  $N=36$ (dashed line) with $J=\varepsilon/2$. Only minor finite-size effects are visible.}
\end{center}
\end{figure}

\begin{figure}
\begin{center}
\includegraphics[scale=0.25,angle=0]{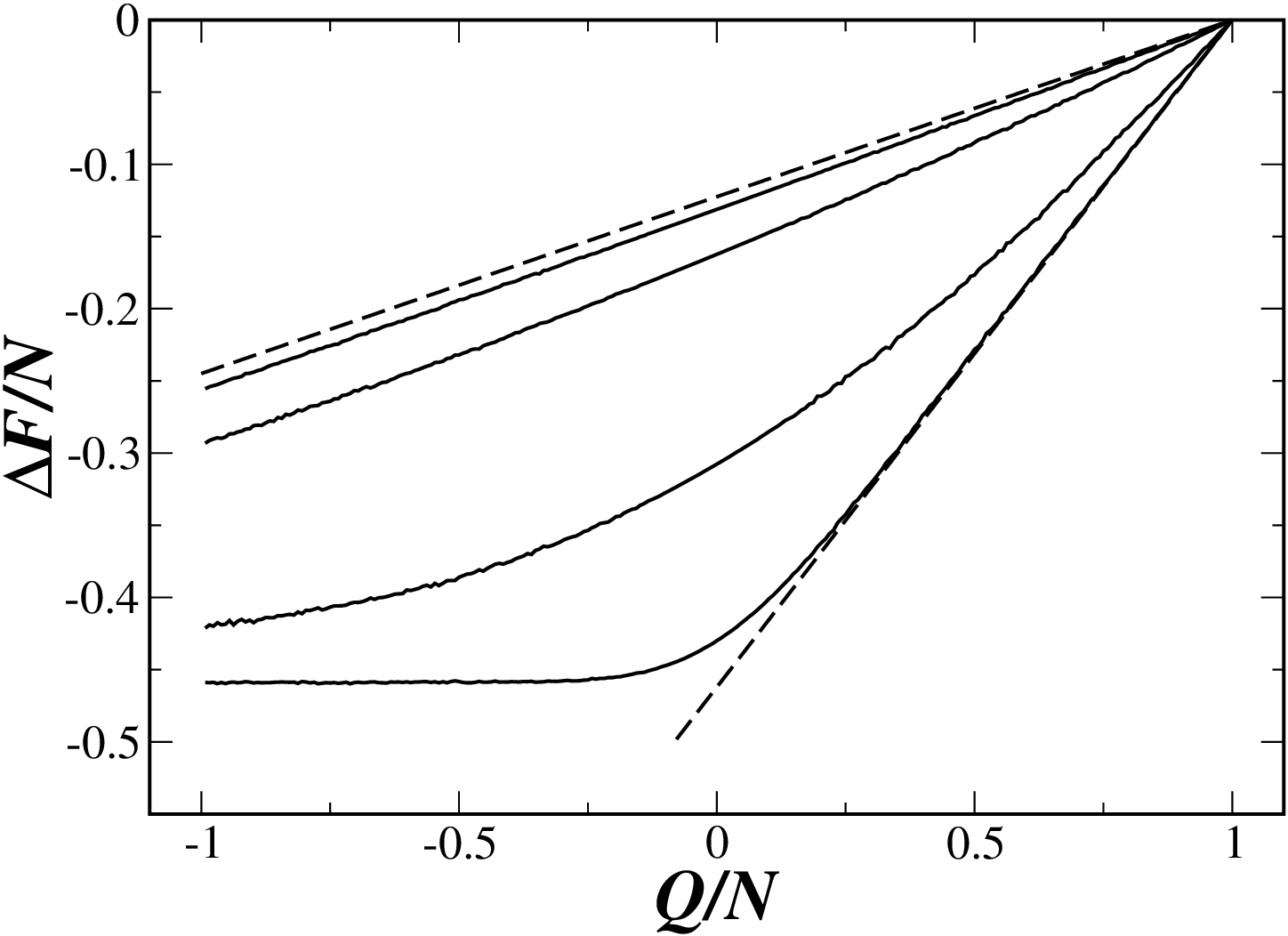}
\caption{\label{bild:coop_freiedifferenz} Free energy difference per site
  (in arbitrary units) of the association of the probe ensemble with the two competing
  molecules as a function of their similarity for different
  cooperativities $J$ (with $\beta_{\mb{\tiny D}}=\beta=0.5$). 
  For the upper dashed line $J=0$, the lower dashed line
  describes the limiting case $J\to \infty$ for $Q/N$ close
  to one \cite{Behringer_etal_vor}. The full curves from top to bottom  
  correspond to the same values of $J$ as in Fig. \ref{bild:coop_comp}.}
\end{center}
\end{figure}

For simplicity, we consider asymptotically large interfaces on a 
square lattice. (The actual calculations shown here were carried out with 
$N=256$, and we checked that the results do not change any more for 
larger $N$). Fig. \ref{bild:coop_comp} shows the average complementarity 
$\left< K \right>/N $ for different cooperativities $J$.
Cooperativity is found to increase the average complementarity of
the designed probe molecules for large enough values of the parameter
$\varepsilon \beta_{\mb{\tiny D}}$. For 
$\varepsilon \beta_{\mb{\tiny D}} \sim 1$, a
small change in the cooperativity $J$ leads to a large difference
in the average complementarity, \ie, small changes in $J$ can
have a large impact on the recognition process. As  $\varepsilon$ is 
typically of the order of 1 kcal/mole, this regime indeed corresponds to 
physiological conditions for reasonable design temperatures,
$1/\beta_{\mb{\tiny D}} \lesssim 1/\beta_{\mb{\tiny Room}}$.
Fig. \ref{bild:coop_freiedifferenz} shows the free energy difference 
per site $\Delta F(Q)/N$ of the association of probe molecules with 
the target structure and a rival structure, for different values
of the cooperativity constant $J$. Increasing the cooperativity increases 
the free energy difference. Relatively small cooperativities are sufficient
to obtain an effect, and the maximum effect of cooperativity is already
reached for a value $J \simeq \varepsilon$. Thus, we find that 
cooperativity indeed improves the recognition ability as expected
for cooperativity constants $J \simeq \varepsilon$. The above findings
were obtained for large interfaces. Although minor finite-size
effects are visible for interfaces of realistic size (with $N
\sim \mathcal{O}(30)$) the general findings discussed
above still hold qualitatively (compare inset of
Fig. \ref{bild:coop_comp}).   

{ In situations where one molecule is flexible
  conformational changes occur.
  However, cooperativity works on the level of residue interactions
  and thus we
  expect that the favorable effect of cooperativity to
  molecular recognition is not spoilt by the entropic contributions
  due to refolding. This, however, needs further investigation. Note that flexible binding has been addressed recently \cite{Wang_2006}.}

Next, we investigate the role of the interplay of interactions for
molecular recognition. This study is motivated by the observation that
antibody-antigen interfaces have very specific properties.
Mutagenesis studies have revealed that the structural interface in
these complexes is different from the functional recognition site made
up of those residues that contribute to the binding energy. Only
approximately one quarter of the residues at the interface contribute
considerably to the binding energy \cite{Cunningham_1993,
  LoConte_1999}. These contributing residues are sometimes called
``hot spots''. In addition it has been shown that antigen-antibody
interfaces are less hydrophobic, compared to other protein-protein
interfaces, so that the relatively strong hydrogen bonds are
more important \cite{LoConte_1999}.  In the immune system molecular
recognition must satisfy very specific requirements. The immune system
has to recognize substances that have never been encountered before.
Thus antigen-antibody recognition has to exhibit a large flexibility
\cite{Jones_1996}, and has to be able to adapt very rapidly by
evolution. These peculiarities of antibody-antigen interfaces suggest
that selective molecular interactions are obtained most efficiently
with only a few strong interactions across the interface, so that a
complementarity with the whole recognition site is not necessary.

Within our two-stage approach, we can address the question whether 
few but strong bonds or many but weak bonds are more favorable.
To this end we consider a model which distinguishes between active
and inactive residues only. Only active residues contribute to a bond.
The variables $\sigma$ and $\theta$ now take on the values
$\sigma_i,\theta_i = +1$ for active and $\sigma_i,\theta_i = 0$ for 
inactive residues, and the Hamiltonian is given by
\begin{equation}
\label{eq:AI-hamiltonian}
  \mathcal{H}(\sigma,\theta, S) =- \varepsilon_{\sts \mb{H}} 
\sum_{i=1}^N\frac{1+S_i}{2} \sigma_i\theta_i
\end{equation}  
with $S_i$ specifying again the quality of the contact of residues, and
$\varepsilon_{\mb{\tiny H}}$ giving the interaction strength. Moreover,
we extend the design step by fixing the average number of active residues 
$A = \langle \sum_i \theta_i \rangle$ on the probe molecules with a
Lagrange parameter. The total interaction energy $E$ is also subject to 
restrictions: It has to exceed the thermal energy to  
stabilize the complex, but on the other hand it has to be small enough to ensure 
the high flexibility of the target-probe complex that is crucial for the 
immune system. When increasing the average number of active residues $A$, one
must therefore reduce the interaction energy $\varepsilon_{\sts \mb{H}}$
accordingly, \eg, by keeping the product 
$E \approx  A  \varepsilon_{\sts \mb{H}}$ constant.

\begin{figure}
\begin{center}
\includegraphics[scale=0.25,angle=0]{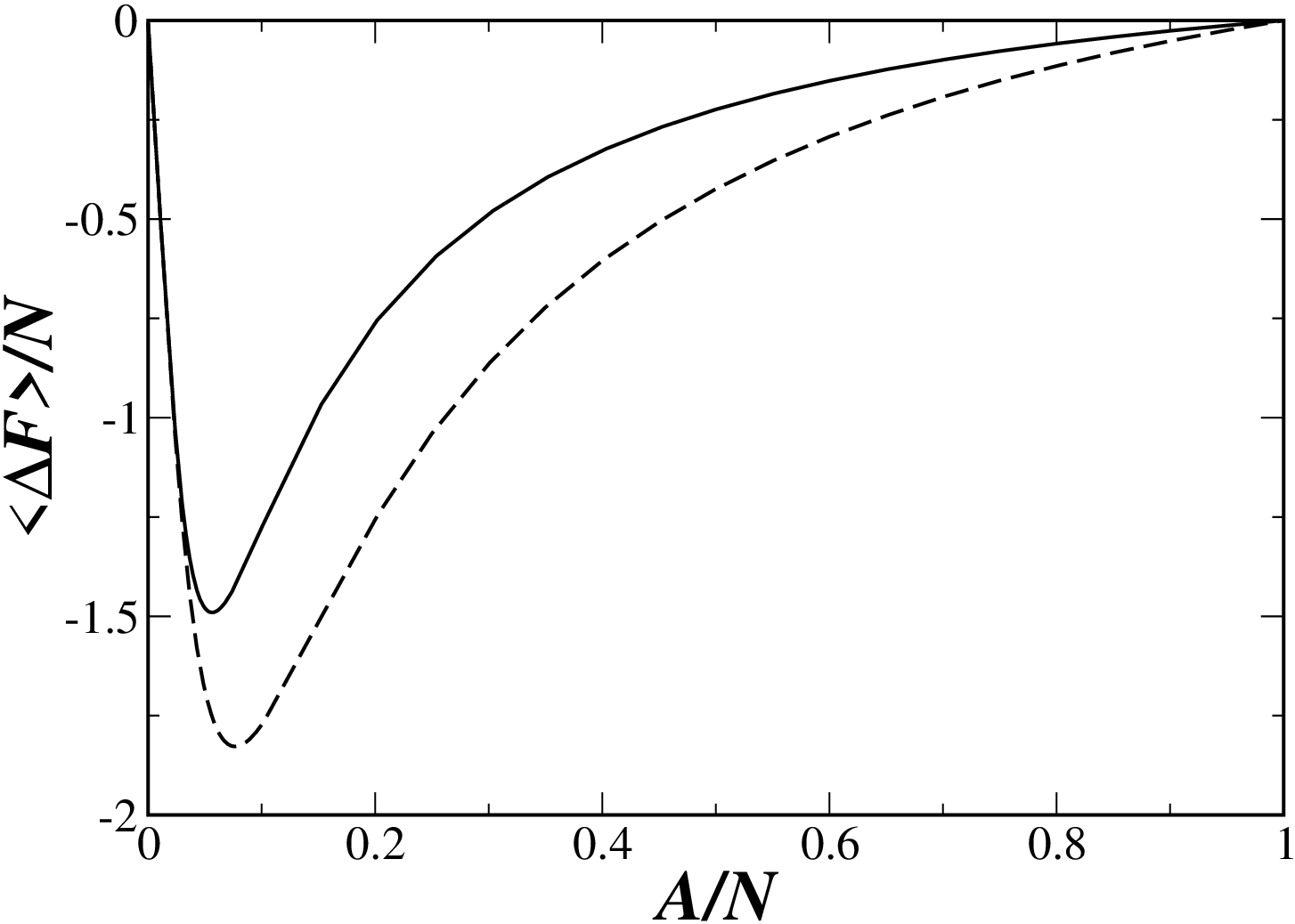}
\caption{\label{bild:lern_NBenergie} Averaged free energy difference
  per site (in arbitrary units)
  as a function of the fraction $A/N$ of active residues for
  $\varepsilon_{\sts \mb{H}} A/N = 0.1$. The full curve corresponds to
  a ratio $\beta/\beta_{\mb{\tiny D}} = 1$, the dashed curve to $\beta/\beta_{\mb{\tiny D}} = 1/2$.
  }
\end{center}
\end{figure}

Figure \ref{bild:lern_NBenergie} shows as a function of $A/N$ the average
free energy difference per site $\Delta F/N$ of the association 
with the target molecule and a rival molecule, averaged over all possible
target and rival structures $\sigma$. We find that $\langle \Delta F \rangle$
exhibits a minimum at a small fraction $A/N$ of active residues. The
position of the minimum at small fractions of $A/N$ is fairly
insensitive to a variation of the interaction parameters. 
Hence this simple coarse-grained model already predicts that molecular 
recognition is most efficient if the functional recognition site  consists 
only of a small fraction of the structural recognition site, as is
indeed observed in antibody-antigen complexes.

In conclusion, we have presented coarse-grained models which
allow to study generic features of biomolecular recognition. 
A two-stage approach which distinguishes between the design of probe
molecules and the test of their recognition abilities has been adopted.
We have applied the approach to investigate the role of cooperativity and
of hydrogen bonding for molecular recognition. It turned out
that cooperativity can substantially influence the efficiency of both
design and recognition ability of recognition sites. Our model also 
reproduces the observation that the structural recognition site has to be
distinguished from a functional recognition site in highly flexible
complexes such as antigen-antibody complexes. 

The approach can readily be generalized to study other aspects of
molecular recognition. For example, it will be interesting to investigate
the influence of the heterogeneity of the mixture of target and rival 
molecules in physiological situations. This can be incorporated by considering 
ensembles of targets and rivals differing in certain properties as for
example correlations and length scales. A recent study indeed showed 
that the local small-scale structure of molecules seems to be important 
for molecular recognition \cite{Bogner_2004}.

Financial support of the Deutsche Forschungsgemeinschaft (SFB 613) is gratefully acknowledged.


\begin{thebibliography}{66}

\bibitem{Alberts_1994} B. Alberts et. al, {\it Molecular Biology of the Cell}, Garland
  Publishing, Inc., New York, 1994.



\bibitem{Kleanthous_2000} C. Kleanthous, ed., {\it Protein-Protein
    Recognition}, Oxford University Press, Oxford, 2000.

\bibitem{pauling_1940} L. Pauling, M. Delbr\"uck, Science {\bf 92}, 77 (1940).


\bibitem{Chakraborty_2001} A. K. Chakraborty, Phys. Rep. {\bf 342}, 1 (2001).

\bibitem{Polotsky_2004a} A. Polotsky, A. Degenhard, F. Schmid,
  J. Chem. Phys. {\bf 120}, 6246 (2004); {\bf 121}, 4853 (2004).


\bibitem{Bogner_2004} T. Bogner, A. Degenhard, F. Schmid, 
  Phys. Rev. Lett. {\bf 93}, 268108 (2004).

\bibitem{Janin_1986} J. Janin, 
 Proteins {\bf 25}, 438 (1986).

\bibitem{Wang_2003} J. Wang, G. M. Verkhivker, 
  Phys. Rev. Lett. {\bf 90}, 188101 (2003).

\bibitem{Jones_1996} S. Jones, J. M. Thornton, 
  Proc. Natl. Acad. Sci. USA, {\bf 93}, 13 (1996).

\bibitem{LoConte_1999} L.\,Lo Conte, C.\,Chothia, J.\,Janin, 
  {J. Mol. Biol.} {\bf 285}, 2177 (1999). 


\bibitem{Pande_2000} V. S. Pande, A. Yu. Grosberg, T. Tanaka,
  {Rev. Mod. Phys.} {\bf 72}, 259 (2000).

\bibitem{Jayaraman_2005} A. Jayaraman, C. K. Hall, J. Genzer, 
  {Phys. Rev. Lett.} {\bf 94}, 078103 (2005).

\bibitem{Dill_1985} K. A. Dill, 
  {Biochemistry} {\bf 24}, 1501 (1985).

\bibitem{footnote}
Note that the original HP-model does not contain an additional 
variable $S$ to model the quality of contacts.

\bibitem{Cera_1998} E. di Cera, 
  {Chem. Rev.} {\bf 98}, 1563 (1998).

\bibitem{Hueller_2002} A. H\"uller, M. Pleimling, 
  {Int. J. Mod. Phys. C} {\bf 13}, 947 (2002); F. Wang, D. P. Landau, 
 {Phys. Rev. Lett.} {\bf 86}, 2050 (2001)


\bibitem{Wang_2006} J. Wang, Q. Lu, H. P. Lu, PLoS Comput. Biol. {\bf
    2}, e78 (2006).


\bibitem{Behringer_etal_vor} H. Behringer, A. Degenhard, F. Schmid,
  unpublished.
  
\bibitem{Cunningham_1993} B.\,C.\,Cunningham, J.\,A.\,Wells, {J.  Mol. Biol.} 
  {\bf 234}, 554 (1993).

\end{thebibliography}
\end{document}